\documentclass{JHEP3}

\usepackage{amssymb}
\usepackage{url}
\usepackage{ifthen,graphics,epsfig}

\newcommand{\eV}{{e\kern-.07em V}}

\makeatletter
\newdimen\z@ \z@=0pt 
\newskip\z@skip \z@skip=0pt plus0pt minus0pt
\def\m@th{\mathsurround=\z@}
\def\ialign{\everycr{}\tabskip\z@skip\halign} 
\def\eqalign#1{\null\,\vcenter{\openup\jot\m@th
  \ialign{\strut\hfil$\displaystyle{##}$&$\displaystyle{{}##}$\hfil
      \crcr#1\crcr}}\,}
\makeatother

\title{A global fit to determine the pseudoscalar mixing angle and the gluonium content of the $\eta$' meson}

\author{The KLOE collaboration:\\
F.~Ambrosino,$^{c,d}$
A.~Antonelli,$^a$
M.~Antonelli,$^a$
F.~Archilli,$^{h,i}$
P.~Beltrame,$^b$
G.~Bencivenni,$^a$
S.~Bertolucci,$^a$
C.~Bini,$^{f,g}$
C.~Bloise,$^a$
S.~Bocchetta,$^{j,k}$
F.~Bossi,$^a$
P.~Branchini,$^k$
G.~Capon,$^a$
T.~Capussela,$^a$
F.~Ceradini,$^{j,k}$
P.~Ciambrone,$^a$
E.~De~Lucia,$^a$
A.~De~Santis,$^{f,g}$
P.~De~Simone,$^a$
G.~De~Zorzi,$^{f,g}$
A.~Denig,$^b$
A.~Di~Domenico,$^{f,g}$
C.~Di~Donato,$^d$
B.~Di~Micco,$^{j,k}$$^1$~
M.~Dreucci,$^a$
G.~Felici,$^a$
S.~Fiore,$^{f,g}$
P.~Franzini,$^{f,g}$
C.~Gatti,$^a$
P.~Gauzzi,$^{f,g}$
S.~Giovannella,$^a$
E.~Graziani,$^k$
M.~Jacewicz,$^a$
G.~Lanfranchi,$^a$
J.~Lee-Franzini,$^{a,\ell}$
M.~Martini,$^{a,e}$
P.~Massarotti,$^{c,d}$
S.~Meola,$^{c,d}$
S.~Miscetti,$^a$
M.~Moulson,$^a$
S.~M\"uller,$^b$
F.~Murtas,$^a$
M.~Napolitano,$^{c,d}$
F.~Nguyen,$^{j,k}$
M.~Palutan,$^a$
E.~Pasqualucci,$^g$
A.~Passeri,$^k$
V.~Patera,$^{a,e}$
P.~Santangelo,$^a$
B.~Sciascia,$^a$
T.~Spadaro,$^a$
M.~Testa,$^{f,g}$
L.~Tortora,$^k$
P.~Valente,$^g$
G.~Venanzoni,$^a$
R.Versaci,$^{a,e}$
G.~Xu,$^{a,m}$
\\
\llap{$^a$}Laboratori Nazionali di Frascati dell'INFN, Frascati, Italy\\
\llap{$^b$}Institut f\"ur Kernphysik, Johannes Gutenberg - Universit\"at Mainz, Germany\\
\llap{$^c$}Dipartimento di Scienze Fisiche dell'Universit\`a  ``Federico II'', Napoli, Italy\\
\llap{$^d$}INFN Sezione di Napoli, Napoli, Italy\\
\llap{$^e$}Dipartimento di Energetica dell'Universit\`a ``La Sapienza'', Roma, Italy\\
\llap{$^f$}Dipartimento di Fisica dell'Universit\`a ``La Sapienza'' , Roma, Italy\\
\llap{$^g$}INFN Sezione di Roma, Roma, Italy\\ 
\llap{$^h$}Dipartimento di Fisica dell'Universit\`a ``Tor Vergata'', Roma, Italy\\
\llap{$^i$}INFN Sezione di Roma Tor Vergata, Roma, Italy\\  
\llap{$^j$}Dipartimento di Fisica dell'Universit\`a ``Roma Tre'', Roma, Italy\\
\llap{$^k$}INFN Sezione di Roma Tre, Roma, Italy\\  
\llap{$^\ell$}Physics Department, State University of New York at Stony Brook, USA\\
\llap{$^i$}Institute of High Energy Physics of Academia Sinica,  Beijing, China\\

}
\footnotetext[1]{
Corresponding authors:\makebox{~B.~Di~Micco~\email{dimicco@fis.uniroma3.it}}}

\abstract{
We update the values of the $\eta$-$\eta'$ mixing angle and  of the $\eta'$ gluonium content by fitting our measurement $R_{\phi} = {\rm BR}(\phi \to \eta' \gamma)/{\rm BR}(\phi  \to \eta \gamma)$ together with several vector meson  radiative decays to pseudoscalars  $(V \to P \gamma)$, pseudoscalar mesons radiative decays to vectors  $(P \to V \gamma)$  and the $\eta' \to \gamma \gamma$, $\pi^0 \to \gamma \gamma$ widths. From the fit we extract a gluonium fraction of $Z^2_G =  0.12 \pm 0.04$, the pseudoscalar mixing angle $\psi_P = (40.4 \pm 0.6)^\circ$ and the $\phi-\omega$ mixing angle 
$\psi_V = (3.32 \pm 0.09)^\circ $. $Z^2_G$ and $\psi_P$ are fairly consistent with those previously published. We also evaluate the impact on the $\eta'$ gluonium content determination of future experimental improvements of the $\eta'$ branching ratios and decay width. }

\received{\today}
\accepted{\today}
\preprint{}
\begin{document}

\overfullrule10pt

\section{Introduction}
The $\eta'$ meson, being almost a pure SU(3)$_{\rm   flavour}$ singlet,
is considered a good candidate to host a gluon condensate. The question of a gluonium component in the $\eta'$ meson has been extensively investigated  in the past but  it is still without a definitive conclusion \cite{Referenze}. We extract the $\eta'$ gluonium content and the $\eta$-$\eta'$ mixing angle in the constituent quark model 
according to the  Rosner \cite{Rosner} approach with the modifications introduced in ref. \cite{ESCRZ} as described in the following.
We use the same method of ref. \cite{Escribano_new}; in addition, we  also introduce in the fit the $\pi^0 \to \gamma \gamma$ 
and $\eta' \to \gamma \gamma$ branching fractions according to the prescription of ref. \cite{Kou}.  This method relates our measurement of the ratio $\phi \to \eta' \gamma$ and $\phi \to \eta \gamma$ branching ratio (BR), $R_{\phi} = {\rm BR}(\phi \to \eta' \gamma)/{\rm BR}(\phi \to \eta \gamma)$ \cite{rphipaper}, to the $\eta'$ gluonium content and to the $\eta, \eta'$ mixing angle. The same quantities were extracted in our previous analysis \cite{rphipaper} with some assumptions. This has given rise to some objections from refs. \cite{Escribano_new} and \cite{Thomas}. Here we give an answer to these objections and we repeat the fit taking into account their comments. Then we repeat the fit with recently updated experimental results.
  
The $\eta$ and $\eta'$ states can be represented in the base  $\left |N \right \rangle=(\left |u\bar u \right \rangle+\left |d\bar d\right \rangle)/\sqrt2$, $\left |S\right \rangle = |s\bar s\rangle$ and $\left |G \right \rangle=\left |{\rm gluonium} \right \rangle$ as:  
\begin{eqnarray}
|\eta'\rangle&=&\cos\psi_G\,\sin\psi_P\,|N\rangle+\cos\psi_G\,\cos\psi_P\,|S\rangle + \sin\psi_G\,|G\rangle\\
|\eta\rangle& = &\cos\psi_P|N\rangle-\sin\psi_P|S\rangle\label{eq:etaquark}
\end{eqnarray}
where $\psi_P$ is the $\eta$-$\eta'$ mixing angle and $Z^2_G = \mathrm{sin}^2 \psi_G$ is the gluonium fraction in the $\eta'$ meson. According to ref. \cite{Cheng} the state $|G\rangle$ could be the $\eta$(1405): a pseudoscalar glue ball candidate.

The ratio $R_{\phi } = {\rm BR}(\phi  \to \eta' \gamma)/{\rm BR}(\phi  \to \eta \gamma)$
is related to the $\psi_P$ and $\psi_G$ parameters by the formula \cite{rphipaper} :
\begin{equation}
\label{eq:mixing}
R_{\phi } = \mathrm{cot}^{2}\varphi_{P} \mathrm{cos}^2 \varphi_{G} \left(
1-\frac{m_s}{{\bar m}}\frac{Z_{q}}{Z_{s}}\frac{\mathrm{tan} \psi_V}{\mathrm{sin}2\varphi_{P}}\right )^2
\left( \frac{p_{\eta^{\prime}}}{p_{\eta}}  \right )^3
\end{equation}
where $p_{\eta'}$ and $p_{\eta}$ are the momenta of the $\eta'$ and $\eta$ meson respectively in the $\phi$ reference frame, $m_{s}/\bar{m} = 2m_{s}/(m_u+m_d)$ is the constituent quark masses ratio and $\psi_V$ is the $\phi$-$\omega$  mixing angle. Following ref. \cite{ESCRZ} we define the constant $C_{q} = \left < q \bar{q}_\rho| q\bar{q}_\eta \right >$ as the overlap between the spatial wave functions of the quark-antiquark pair in the $\rho$ and the $\eta$ meson. Isospin symmetry is assumed exact, so that $m_u = m_d = \bar{m}$ and the following further relations follow:
\[
C_q = \left <q\bar{q}_{\eta} \right | \left . q\bar{q}_\omega \right > = \left <q\bar{q}_\eta \right | \left . q\bar{q}_\rho \right >, \quad C_s = \left <s\bar{s}_\eta \right | \left . s \bar{s}_\phi \right >, \quad C_\pi = \left <q\bar{q}_\pi \right | \left . q\bar{q}_\omega \right > = \left <q\bar{q}_\pi \right | \left . q\bar{q}_\rho \right >
\]
where we indicate with $\left | q\bar{q}_\eta \right >$ and $\left | q\bar{q}_\omega \right >$ the $q \bar{q}$ spatial wave function in the $\eta$ and $\omega$ mesons, 
and with  $|s\bar{s}_\eta\rangle$ and $|s\bar{s}_\phi\rangle$  the $\bar{s}s$ spatial wave function in the $\eta$ and $\phi$ mesons.
The parameters $Z_{q}$ and $Z_s$ are the ratios: $Z_{q} = C_q/C_\pi$ and $Z_s = C_s/C_\pi$. In this model SU(3)$_{\rm flavour}$ breaking effects are accounted for by the different values of the effective quark masses, $m_s > m_u = m_d = \bar{m}$, and
by $Z_q \neq Z_s$.  

 In our previous analysis \cite{rphipaper} the parameters $Z_s$, $Z_{q}$, $\psi_V$ and $m_s/\bar{m}$  were taken from ref. \cite{ESCRZ} where  BR($\phi \to \eta' \gamma$) and BR($\phi  \to \eta \gamma$) were fitted together with other  $V \to P \gamma$ decay rates ($V$ indicates the vector mesons $\rho,\omega,\phi $ and $P$ the pseudoscalars $\pi^0, \eta, \eta'$) assuming no $\eta'$ gluonium content. We fitted \cite{rphipaper} our measurement 
\[
R_{\phi } = \frac{{\rm BR}(\phi  \to \eta' \gamma)}{{\rm BR}(\phi  \to \eta \gamma)} = (4.77 \pm 0.09_{stat.} \pm 0.19_{syst.}) \times 10^{-3}
\]

\noindent together with the available data \cite{PDG06} on $\Gamma(\eta' \to \gamma \gamma)/\Gamma(\pi^0 \to \gamma \gamma)$,
$\Gamma(\eta' \to \rho \gamma)/\Gamma(\omega \to \pi^0 \gamma)$ and $\Gamma(\eta' \to \omega \gamma)/\Gamma(\omega \to \pi^0 \gamma)$. 
The dependence of these ratios from the mixing angle $\psi_P$ and the gluonium content $\psi_G$ is given by the following equations:

\[
X_{\eta'} = \mathrm{sin}\psi_P \, \mathrm{cos}\psi_G, \, Y_{\eta'}= \mathrm{cos}\psi_P \, \mathrm{cos}\psi_G 
\]
\begin{eqnarray}
\frac{\Gamma(\eta' \to \gamma \gamma)}{\Gamma(\pi^0 \to \gamma \gamma)}  & = &
\frac{1}{9} \left( \frac{m_{\eta'}}{m_{\pi^0}} \right)^3 \left(5\frac{f_{\pi}}{f_q} \, \mathrm{cos} \psi_G \, \mathrm{sin} \psi_P + \sqrt{2}\frac{f_\pi}{f_s} \, \mathrm{cos} \psi_G \, \mathrm{cos} \psi_P \right )^2   \label{eq:constrfirst} \\
\frac{\Gamma(\eta' \to \rho \gamma)}{\Gamma(\omega \to \pi^0 \gamma)} & = &
3\frac{Z^2_{q}}{\mathrm{cos}^2(\psi_V)}\left(\frac{m_{\eta'}^2-m_{\rho}^2}{m_{\omega}^2-m_{\pi}^2}\cdot \frac{m_{\omega}}{m_{\eta'}} \right)^3 X^2_{\eta'} \label{eq:etaprhog}  \\ 
\frac{\Gamma(\eta' \to \omega \gamma)}{\Gamma(\omega \to \pi^0 \gamma)}  & = & 
\frac{1}{3}\left(\frac{m_{\eta'}^2-m^2_{\omega}}{m^2_{\omega}-m^2_{\pi}}\cdot \frac{m_{\omega}}{m_{\eta'}}\right)^3
\left[Z_{q} X_{\eta'}+2\frac{\bar{m}}{m_s}Z_s\cdot \mathrm{tan}{\psi_V}Y_{\eta'}\right]^2.   \label{eq:etapomegag} 
\end{eqnarray}
where $f_\pi$ is the pion decay constant and $f_q$ and $f_s$ are the decay constants of the isospin singlet states (mainly $\eta,\eta'$ mesons) at  no anomaly limit \cite{Kou}. 
The fit result was $\psi_P = (39.7 \pm 0.7)^\circ$ and $Z^2_G = \mathrm{sin}^2{\psi_G} = 0.14 \pm 0.04$,  $P(\chi^2) = 49$\%. Imposing $\psi_G = 0$ the $\chi^2$ probability of the fit decreased to 1\%.

In ref. \cite{Escribano_new}, a procedure similar to \cite{ESCRZ} has been adopted, but they fitted also the gluonium component in the $\eta'$ wave function that was previously fixed at zero.  The result $Z^2_G = 0.04 \pm 0.09$ deviates  1$\sigma$ from our value but with a larger error. In refs.\cite{Escribano_new} and \cite{Thomas} this difference was attributed to the use in our fit of the parameters $Z_s$ and $Z_q$ obtained in ref. \cite{ESCRZ} assuming no gluonium content. However further tests of the fit procedure showed that $Z^2_G$ and $\psi_P$ are marginally sensitive to large variations of $Z_q$ and $Z_s$ \cite{HAdron_EPJ}.
Here we repeat the fit with a larger number of free parameters, including $Z_s$ and $Z_q$. 
\section{Fit description}

In order to enlarge the set of parameters used in the fit we add to the equations (\ref{eq:constrfirst}-\ref{eq:etapomegag}) the following further relations that can be derived directly from \cite{Escribano_new}:
\begin{eqnarray}
\frac{\Gamma(\omega \to \eta \gamma)}{\Gamma(\omega \to \pi^0 \gamma)} = & & \frac{1}{9}\left [Z_{q} \, \mathrm{cos}\psi_P\mathrm{} - 2 \frac{\bar{m}}{m_s}Z_s \, \mathrm{tan}\psi_V \, \mathrm{sin}\psi_P \right]^2 \left ( \frac{m^2_{\omega} - m^2_{\eta}}{m^2_{\omega}-m^2_{\pi^0}} \right)^3 \label{eq_first} \\
\frac{\Gamma(\rho \to \eta \gamma)}{\Gamma(\omega \to \pi^0 \gamma)}  = & & Z^2_{q} \frac{\mathrm{cos}^2\psi_P}{\mathrm{cos}^2\psi_V}\left ( \frac{m^2_{\rho} - m^2_{\eta}}{m^2_{\omega}-m^2_{\pi^0}} \frac{m_{\omega}}{m_{\rho}} \right)^3 \\
\frac{\Gamma(\phi \to \eta \gamma)}{\Gamma(\omega \to \pi^0 \gamma)} = & & \frac{1}{9} \left [ Z_{q} \, \mathrm{tan}\psi_V \, \mathrm{cos}\psi_P + 2 \frac{\bar{m}}{m_s} Z_s \, \mathrm{sin}\psi_P\right]^2 \left ( \frac{m^2_{\phi} - m^2_{\eta}}{m^2_{\omega}-m^2_{\pi^0}} \frac{m_{\omega}}{m_{\phi}} \right)^3 \\
\frac{\Gamma(\phi \to \pi^0 \gamma)}{\Gamma(\omega \to \pi^0 \gamma)} = & & \mathrm{tan}^2 \psi_V \cdot \left( \frac{m^2_\phi - m^2_{\pi^0}}{m^2_{\omega} - m^2_{\pi^0}} \frac{m_{\omega}}{m_{\phi}} \right)^3  \label{eq_phiv} \\
\frac{\Gamma(K^{*+} \to K^+ \gamma)}{\Gamma(K^{*0} \to K^0 \gamma)} = & & \left ( \frac{2\frac{m_s}{\bar{m}}-1}{1+\frac{m_s}{\bar{m}}} \right)^2 \cdot \left(\frac{m^2_{K^{*+}}-m^2_{K^+}}{m^2_{K^{*0}} - m^2_{K^0}} \cdot \frac{m_{K^{*0}}}{m_{K^{*+}}} \right)^3 \label{eq_last}
\end{eqnarray}
Notice that, differently from \cite{Escribano_new} where the VP$\gamma$ couplings are fitted, we fit directly ratios of partial decay widths. This allows both to
reduce the parameters involved in the fit (two of them cancel out in the ratios), and to simplify the error treatment using quantities proportional (directly or inversely) to the experimental measurements. As an example, a ratio of $\Gamma's$ is written as:
\[
\frac{\Gamma(\eta' \to \rho \gamma)}{\Gamma(\omega \to \pi^0 \gamma)} = \frac{\mathrm{BR(}\eta' \to \rho \gamma)}{\mathrm{BR(}\omega \to \pi^0 \gamma)} \frac{\Gamma_{\eta'}}{\Gamma_{\omega}}
\]
In this way the correlation matrix among the $\eta'$ branching ratios and the decay widths can be used directly .
The fit is performed minimising the $\chi^2$ function:
\[
\chi^2 = \sum_{i,j} (y_i - y_i^{\rm th}) \left ({\rm \bf V}^{-1} \right)_{ij}  (y_j-y_j^{\rm th})
\]
where $y_i$ are the experimental measurements of the ratios on the left side of equations (\ref{eq:constrfirst} - \ref{eq:etapomegag}) and (\ref{eq_first} - \ref{eq_last}), $y_i^{\rm th}$ are the theoretical predictions on the right side of the same equations and   \textbf{V}$^{-1}$ is the inverse of the covariance matrix. This last quantity is obtained summing the contribution from the experimental error on branching ratios, decay widths and their correlations and the uncertainty from theoretical inputs. The method is the same of our previous fit \cite{rphipaper}, but only the parameters $f_q/f_{\pi} = 1.00 \pm 0.01$ and $f_s/f_{\pi} = 1.4 \pm 0.014$ \cite{Feldmann} are taken as input, while in the past also $Z_s$, $Z_q$, $\psi_V$ and $m_s/\overline{m}$ were fixed. The parameters $f_q/f_{\pi}$ and $f_s/f_{\pi}$ involve only the ratio $\Gamma(\eta' \to \gamma \gamma)/\Gamma(\pi^0 \to \gamma \gamma)$. 
The contribution from theoretical error is evaluated by standard error propagation:
\[
{\rm \bf V^{th}} = {\bf A \,  C \, A}^{\rm T}
\]
where \textbf{C} is the covariance matrix of the uncorrelated parameters $x_1 = f_q/f_{\pi}$ and $x_2 = f_s/f_{\pi}$ and \textbf{A} is:
\[
{(\bf A)}_{ij} = \frac{\partial y_i^{\rm th}}{\partial x_j}
\]  
The covariance matrix \textbf{V} is indeed:
\[
{\rm \bf V }= {\rm \bf V^{exp}} + {\rm \bf V^{th}}
\]
where \textbf{V}$^{\rm exp}$ is the covariance matrix of the data used in the fit. 
Particularly relevant is the correlation between the $\eta'$ branching fractions and the decay widths shown in tab. \ref{tab_etaprimecorr}.
\TABLE[hbt]{
\centering
\begin{tabular}{c|ccccccc}
$\rho \gamma$ & -0.34 & & & & & &  \\
$\pi^0 \pi^0 \eta$ & -0.78 & -0.29 & & & & & \\ 
$\omega \gamma$ & -0.35 & -0.24 & 0.32 & & & & \\
$\gamma \gamma$ & -0.26 & -0.12 & 0.26 & 0.08 & & & \\
$3 \pi^0$ & -0.28 & -0.11 & 0.35 & 0.11 & 0.09 & & \\
$\Gamma_{\eta'}$ & 0.32 & -0.02 & -0.24 & -0.05 & -0.88 & -0.08 &  \\
\hline
  & $\pi^+ \pi^- \eta$ & $\rho \gamma$ & $\pi^0 \pi^0 \eta$ & $\omega \gamma$ & $\gamma \gamma$ & $3 \pi^0$ & 
\end{tabular}

\caption{Correlation matrix of the $\eta'$ branching ratios from PDG-2006  fit \cite{PDG06}.} \label{tab_etaprimecorr}} 

The $\eta'$ width and the $\eta' \to \gamma \gamma$ branching ratio are 88\% correlated. This is because the $\eta'$ width is evaluated dividing the $\Gamma (\eta' \to \gamma \gamma)$,  obtained by measuring the cross section 
$\sigma(e^+ e^- \to \eta' e^+ e^-)$, by the $\eta' \to \gamma \gamma$ branching ratio. The fit results and the correlation matrix are shown in tab. \ref{tab_results_fit_standard}.
\TABLE[t]{
\centering
\begin{tabular}{c}
\begin{tabular}{c|c|c}
& $Z_G$ free & $Z_G=0$ \\
\hline
${\rm \chi^2/ndf \, (CL)}$ & 5/3 (17\%) & 13/4 (1.1\%)\\
\hline
$Z^2_{G}$ & 0.105 $\pm$ 0.037 & 0 fixed \\
$\psi_P$ & (40.7 $\pm$ 0.7)$^\circ$ & $(41.6 \pm 0.5)^{\circ}$\\
$Z_{q}$ & 0.866 $\pm$ 0.025 & 0.863 $\pm$ 0.024 \\
$Z_s$ & 0.79 $\pm$ 0.05 & 0.78 $\pm$ 0.05 \\
$\psi_V$ & (3.15 $\pm$ 0.10)$^\circ$ &  (3.17 $\pm$ 0.10)$^\circ$ \\
$m_s/\bar{m}$ & 1.24 $\pm$ 0.07 &  1.24 $\pm$ 0.07 \\
\hline
\end{tabular} \\
\\
\begin{tabular}{c|cccccc}
$\psi_P$ & -0.513  \\
$Z_{q}$    & 0.003  & 0.041  \\
$Z_s$       & 0.088 & -0.188  & 0.050 &  \\
$\psi_V$  & -0.068 & -0.019 & 0.150 & 0.077\\
$m_s/\bar{m}$ & 0 & 0 & 0 & 0.935 & 0 \\
\hline
    & $Z^2_G$ & $\psi_P$ & $Z_{q}$  & $Z_s$ & $\psi_V$ &  \\
 \end{tabular}\\
\end{tabular}
\caption{Fit results using the PDG-2006 data \cite{PDG06} and their correlation matrix.} \label{tab_results_fit_standard}
}
The gluonium fraction $Z^2_G = 0.105 \pm 0.037$ is   2.8$\sigma$ from zero. In order to identify the measurements which require the presence of the gluonium in the $\eta'$
we have repeated the fit fixing $Z_G$ at zero. The results of the fit are shown in tab. \ref{tab_results_fit_standard}.
The $\chi^2$ probability is now quite low, reflecting the 2.8$\sigma$ effect seen in the previous fit, while the pseudoscalar mixing angle is quite stable.

In fig. \ref{fig_pulls} we show for each measurement the pulls of the two fits, defined as 
$p_i = (y_i - y^{\rm th}_i)/\sigma_{y_{i }}$.
\FIGURE{
 \centering

\includegraphics[width=0.5\textwidth]{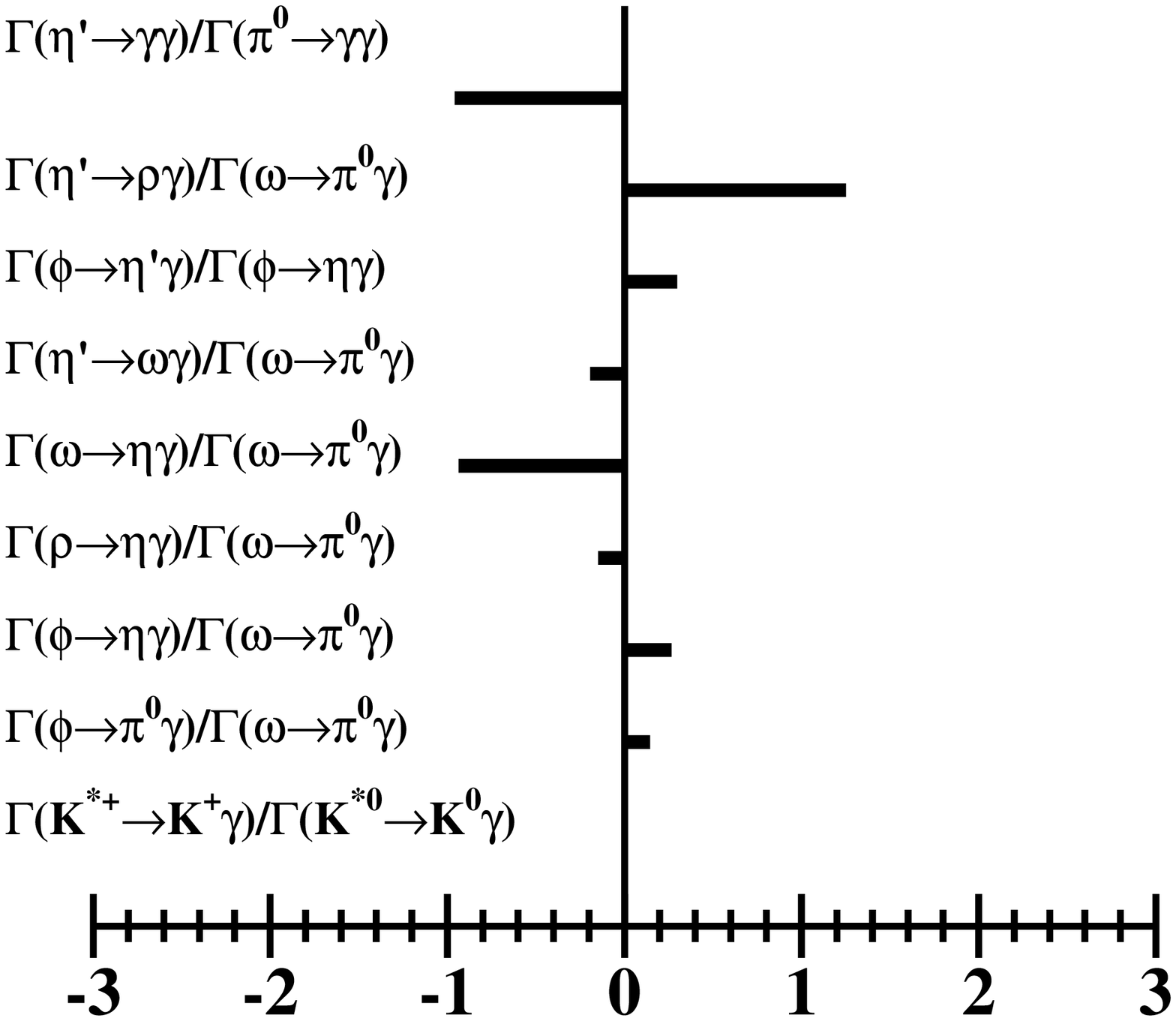}\includegraphics[width=0.5\textwidth]{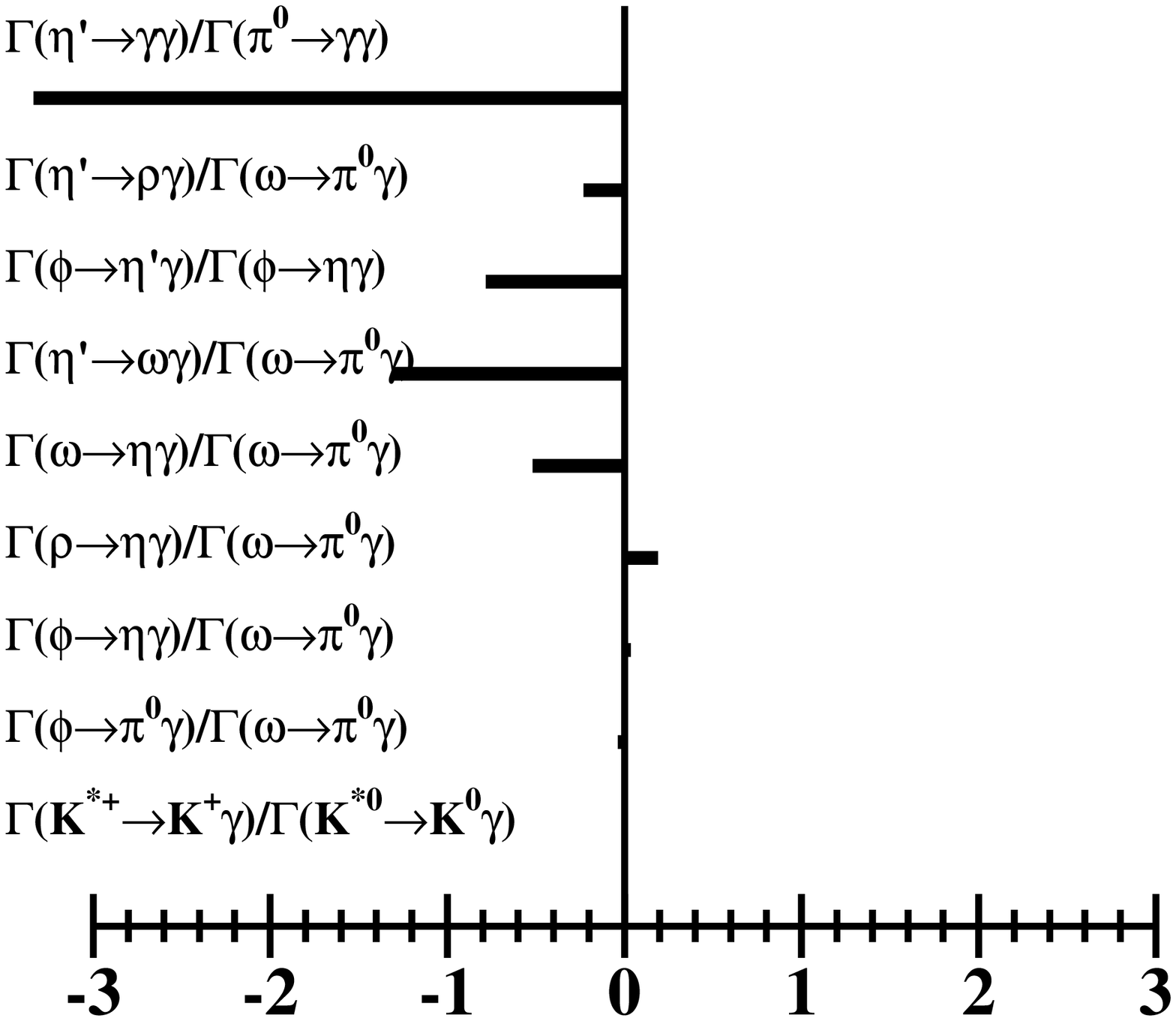}

\caption{Pulls of the fit shown in tab. \ref{tab_results_fit_standard}, {\bf left:} $Z_G$ free, {\bf right:} $Z_G = 0$ (fixed).} \label{fig_pulls}}
The measurement which does not fit in the \emph{no-gluonium} hypothesis is the ratio 
$\Gamma(\eta' \to \gamma \gamma)/\Gamma(\pi^0 \to \gamma \gamma)$, whose pull is less than $-3$, bringing the $\chi^2$ 
probability to 1.1\%. It becomes  $\sim$$-1$ when  $gluonium$ is allowed.  
We have then repeated the fit without using the $\Gamma(\eta' \to \gamma \gamma )/ \Gamma(\pi^0 \to \gamma \gamma)$ information. The result is compared with ref. \cite{Escribano_new} in tab. \ref{tab_escrnoicomparison}.
\TABLE[hbt]{
\begin{tabular}{c|c|c}
& Fit with PDG-2006 & Fit of \\
 &                                   &   ref. \cite{Escribano_new} \\
\hline
${\rm \chi^2/ndf \, (CL)} $ & 1.8/2 (41\%) & 4.2/4 (38\%) \\
\hline
$Z^2_{G}$ & 0.03 $\pm$ 0.06 & 0.04 $\pm$ 0.09 \\
$\psi_G$ & $(10 \pm 10)^\circ$  & $(12 \pm 13)^\circ$ \\
\hline
$\psi_P$ & (41.6 $\pm$ 0.8)$^\circ$ & $(41.4 \pm 1.3)^{\circ}$ \\
$Z_{q}$ & 0.85 $\pm$ 0.03 & 0.86 $\pm$ 0.03 \\
$Z_s$ & 0.78 $\pm$ 0.05 & 0.79 $\pm$ 0.05 \\
$\psi_V$ & (3.16 $\pm$ 0.10)$^\circ$ & (3.2 $\pm$ 0.1)$^\circ$ \\
$m_s/\bar{m}$ & 1.24 $\pm$ 0.07 & 1.24 $\pm$ 0.07\\
\hline
\end{tabular}
\caption{Comparison among the fit results without the $\eta' \to \gamma \gamma / \pi^0 \to \gamma \gamma$ measurements and the results of ref. \cite{Escribano_new}. PDG-2006 data \cite{PDG06} have been used in both fits.} \label{tab_escrnoicomparison}
}
%
The results of the two fits are in agreement, making it evident that the origin of the discrepancy with ref. \cite{Escribano_new} is due to their neglecting  the $\Gamma(\eta' \to \gamma \gamma)/\Gamma(\pi^0 \to \gamma \gamma)$ datum. In ref. \cite{Escribano_new} the couplings among the vectors and the pseudoscalar mesons are used in place  of the width ratio.
The couplings are related to the partial decay width by the following formulae:
\[
\Gamma(V \to P \gamma) = \frac{1}{3}\frac{g^2_{VP\gamma}}{4\pi}|\vec{p}_\gamma|^3, \quad \Gamma(P\to V \gamma) = \frac{g^2_{VP\gamma}}{4\pi}|\vec{p}_{\gamma}|^3
\]
In order to make a full comparison between the two methods we have performed the fit also using the couplings and we have obtain the same results \cite{KLOE_note}.

\section{Update with the recent PDG results}

In the Review of Particle Physics \cite{PDG08} new measurements of the $\rho$, $\omega$, $\eta$ and $\eta'$ mesons have been included,
which change  slightly the partial decay widths used in the fit. Therefore we repeat the fit using these updated values together with our $R_{\phi}$ measurement. All the correlation coefficients among the measurements are taken into account in the fit.
The results of the fit are shown in tab. \ref{tab_fitpdg08} and the correlation matrix in tab. \ref{tab_corrfitpdg08}; the pulls of the fit are shown in fig. \ref{fig_pulls_pdg08}. 

\begin{table}
\centering
\begin{tabular}{c|c|c}
& $Z_G$ free & $Z_G = 0$ fixed \\
\hline
${\rm \chi^2/ndf \, (CL)} $ & 7.9/3 (5\%) & 15/4 ($5\times10^{-3}$)\\
\hline
$Z^2_{G}$ & 0.097 $\pm$ 0.037 & 0 fixed \\
$\psi_P$ & (41.0 $\pm$ 0.7)$^\circ$ & $(41.7 \pm 0.5)^{\circ}$\\
$Z_{q}$ & 0.86 $\pm$ 0.02 & 0.86 $\pm$ 0.02 \\
$Z_s$ & 0.79 $\pm$ 0.05 & 0.78 $\pm$ 0.05 \\
$\psi_V$ & (3.17 $\pm$ 0.09)$^\circ$ &  (3.19 $\pm$ 0.09)$^\circ$ \\
$m_s/\bar{m}$ & 1.24 $\pm$ 0.07 &  1.24 $\pm$ 0.07 \\
\hline
\end{tabular}
\caption{Fit results using the PDG-2008 data.} \label{tab_fitpdg08}
\end{table}
\TABLE{
\centering
\begin{tabular}{c|cccccc}
$\psi_P$ & -0.502  &   &   &  & & \\
$Z_{q}$    & -0.072  & 0.161 & & & & \\
$Z_s$       & 0.081 & -0.180  & 0.028 & & &  \\
$\psi_V$  & -0.082 & 0.013 & 0.169 & 0.078 & & \\
$m_s/\bar{m}$ & 0 & 0 & 0 & 0.940 & 0 & \\
\hline
    & $Z^2_G$ & $\psi_P$ & $Z_{q}$  & $Z_s$ & $\psi_V$ &  \\
\end{tabular}
\caption{Correlation matrix from the fit shown in tab. \ref{tab_fitpdg08}.} \label{tab_corrfitpdg08}}

\FIGURE{
\centering
\includegraphics[width=0.5\textwidth]{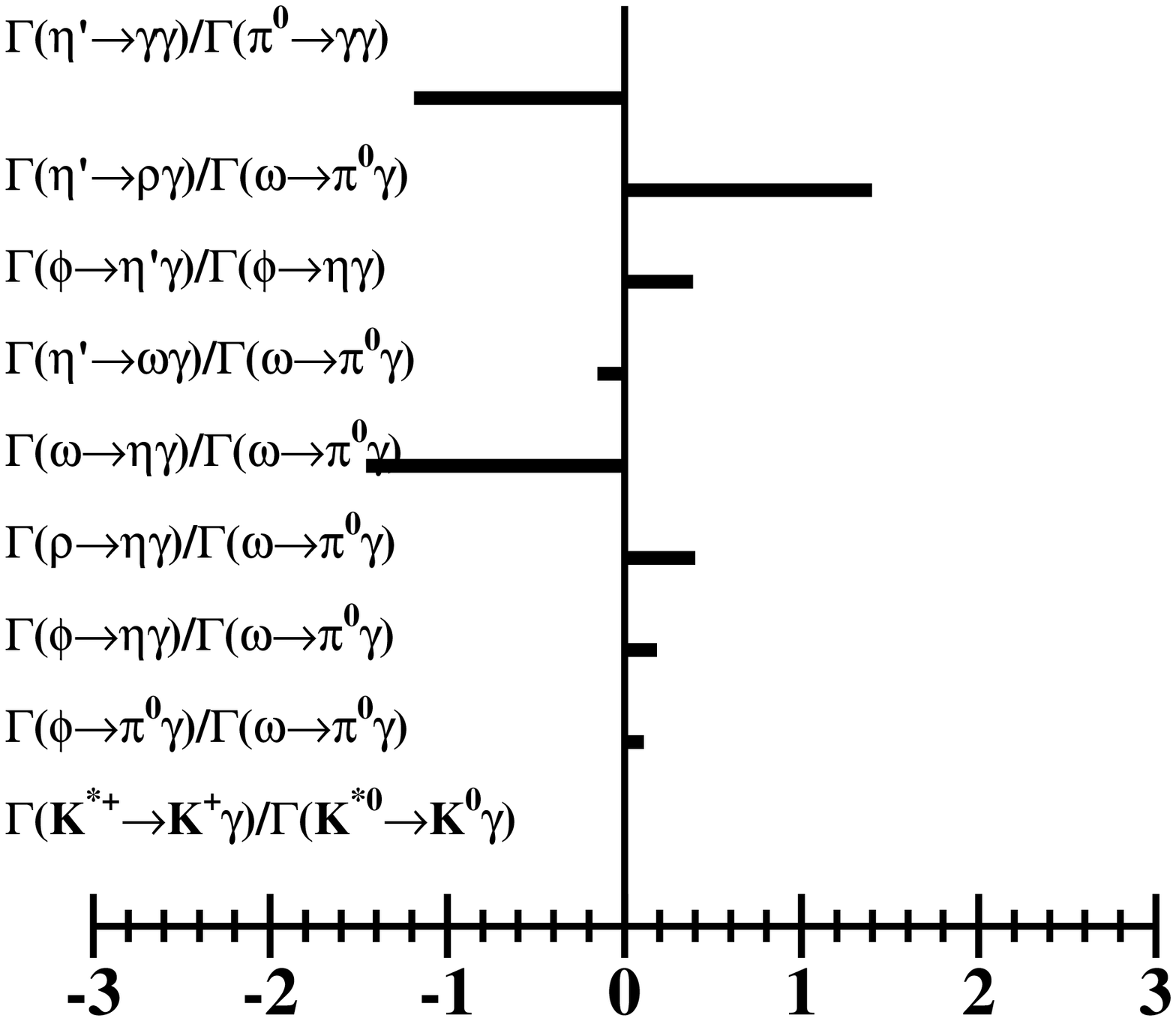}\includegraphics[width=0.5\textwidth]{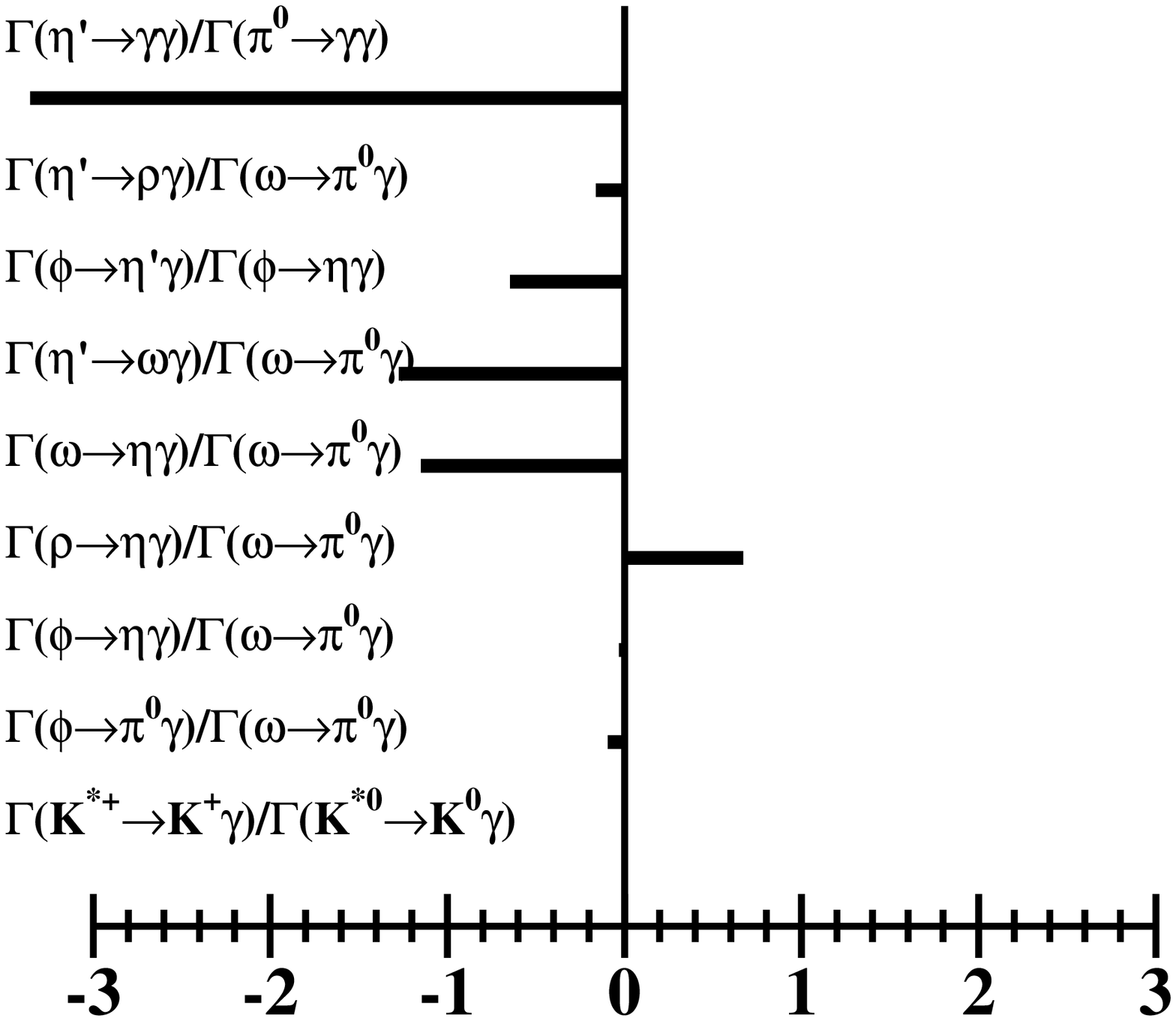}

\caption{Pulls of the fit using PDG-2008 data, 
{\bf left:} $Z_G$ free, {\bf right:} $Z_G=0$  (fixed).} \label{fig_pulls_pdg08}}

The results in tab. \ref{tab_fitpdg08} show that the gluonium hypothesis
is still highly favoured with respect to the null gluonium hypothesis. Nevertheless
the fit probability is quite low also in the gluonium hypothesis: it goes from 17\% using PDG06 data to 5\% using PDG08 data. The reason of the
worsening of the fit is found comparing the pulls of the new fit (fig. \ref{fig_pulls_pdg08}) with the previous one (fig. \ref{fig_pulls}). In particular, the pull of the ratio $\Gamma(\omega \to \eta \gamma)/\Gamma(\omega \to \pi^0 \gamma)$ goes from -0.93 using PDG-2006 data to -1.5 using  
PDG-2008 while the pull of the ratio  $\Gamma(\rho \to \eta \gamma)/\Gamma(\omega \to \pi^0 \gamma)$ goes from -0.14 to +0.39. This happens  because the PDG estimate of the BR$(\omega \to \eta \gamma)$ has changed from $(4.9 \pm 0.5) \times 10^{-4}$ to $(4.6 \pm 0.4) \times 10^{-4}$,  lowering the $\Gamma$'s ratio and worsening the pull\footnote{A fit without the $\Gamma(\eta' \to \gamma \gamma)$/$\Gamma(\pi^0 \to \gamma \gamma)$ ratio has been performed in order to check its effect on the $\omega \to \eta \gamma$ and $\rho \to \eta \gamma$ pulls. In this case the ${\rm \chi^2/ndf}$ of the fit  is high also in the null gluonium hypothesis, nevertheless the $\Gamma(\omega \to \eta \gamma)/\Gamma(\omega \to \pi^0 \gamma)$ pull is -0.68 while the $\Gamma(\rho \to \eta \gamma)/\Gamma(\omega \to \pi^0 \gamma)$ pull is +1.1. In other words the BR($\omega \to \eta \gamma$) fits better  while the BR($\rho \to \eta \gamma$) fits worse. The difference of the two measurements from the best fit is still $\sim$ 2$\sigma$, therefore the poor $\chi^2$ with PDG-2008 data is not due to $\eta' \to \gamma \gamma$ but to the inconsistency between $\Gamma(\omega \to \eta \gamma)$ and $\Gamma(\rho \to \eta \gamma)$ measurements.}.

The BR$(\omega \to \eta \gamma)$ and BR$(\rho \to \eta \gamma)$ PDG  values are dominated by the measurement of the $e^+ e^- \to \eta \gamma$ cross section by SND \cite{SND07}
as a function of $\sqrt{s}$  in the $\rho, \omega, \phi$ mass range. From the measured cross section they extract the $\rho$, $\omega$ partial decay widths assuming the Vector Meson Dominance model and a parametrisation for the $\rho'$ resonance. Some correlation is therefore expected between the $\rho$ and the $\omega$ partial decay widths which are not discussed in ref. \cite{SND07}, moreover the decay widths are model dependent. The average value reported by PDG-2008, BR($\omega \to \eta \gamma$) = $(6.3 \pm 1.3) \times 10^{-4}$, is dominated by a model independent measurement\cite{CBarromegaetag} and is 1.2$\sigma$ away from the PDG fit. Using this value for BR$(\omega \to \eta \gamma)$ we obtain a much better $\chi^2$ probability: $P(\chi^2) = 28\%$  in the gluonium hypothesis  and $1.1\%$ fixing the gluonium at zero \cite{KLOE_note}. Both gluonium content and pseudoscalar  mixing angle are unchanged ($Z^2_G = 0.11 \pm 0.04$, $\psi_P = (40.6 \pm 0.7) ^\circ$ in the gluonium hypothesis). Therefore we will use the average value for  BR$(\omega \to \eta \gamma)$  in the following.

\section{Update with the new KLOE measurement of BR$(\omega \to \pi^0 \gamma)$}

The relations (\ref{eq:etaprhog}-\ref{eq:etapomegag}) and (\ref{eq_first}-\ref{eq_phiv}) are dependent from the $\omega \to \pi^0 \gamma$ decay rate. Recently we have improved the measurement of this branching fraction BR$(\omega \to \pi^0 \gamma) = (8.09 \pm 0.14)$\%  \cite{KLOEompg}. 
This value is about 3$\sigma$ different from the PDG 2008 value:  BR($\omega \to \pi^0 \gamma\mathrm{)} = (8.92 \pm 0.24)$\%. 
We then performed the fit using our measurement of BR($\omega \to \pi^0 \gamma$).

Moreover $f_q/f_{\pi}$ and $f_s/f_{\pi}$ have been fixed according to ref. \cite{Kou}.  In the exact isospin symmetry approximation the relations
\[
f_q = f_{\pi}; \qquad f_s = \sqrt{2f_K^2-f_{\pi}^2}
\]
hold, where $f_{\pi}$ and $f_K$ are the $\pi$ and $K$ decay constants. 
Therefore $f_s/f_{\pi} = \sqrt{2f_K^2/f_{\pi}^2-1}$. 
Using $f_K/f_{\pi}$ from lattice calculation \cite{UKQCD} we get:
\begin{equation} \label{eq:fqfs}
\frac{f_q}{f_{\pi}} = 1 \qquad \frac{f_s}{f_{\pi}} = 1.352 \pm 0.007.
\end{equation}

The results of the fit are shown in  tab. \ref{tab_result_final} and the correlation matrix in tab.\ref{tab_corr_result_final}.
\begin{table}[hbt]
\centering
\begin{tabular}{c|c|c}
& $Z_G$ free & $Z_G=0$ fixed \\
\hline

${\rm \chi^2/ndf \, (CL)} $ & 4.6/3 (20\%) & 14.7/4 (0.5\%)\\
\hline
$Z^2_{G}$ & 0.115 $\pm$ 0.036 & 0 \\
$\psi_P$ & (40.4 $\pm$ 0.6)$^\circ$ & $(41.4 \pm 0.5)^{\circ}$\\
$Z_{q}$ & 0.936 $\pm$ 0.025 & 0.927 $\pm$ 0.023 \\
$Z_s$ & 0.83 $\pm$ 0.05 & 0.82 $\pm$ 0.05 \\
$\psi_V$ & (3.32 $\pm$ 0.09)$^\circ$ &  (3.34 $\pm$ 0.09)$^\circ$ \\
$m_s/\bar{m}$ & 1.24 $\pm$ 0.07 &  1.24 $\pm$ 0.07 \\
\hline
\end{tabular}
\caption{Fit results using PDG-2008 inputs, BR($\omega \to \eta \gamma$) from PDG direct measurement average and the KLOE BR($\omega \to \pi^0 \gamma$) and $R_{\phi}$. The equations (4.1) have been used for the $f_q/f_\pi$ and $f_s/f_\pi$ parameters. } \label{tab_result_final}
\end{table}

\begin{table}
\centering
\begin{tabular}{c|cccccc}

$\psi_P$ & -0.507  \\
$Z_{q}$    & 0.063  & -0.018  \\
$Z_s$       & 0.092 & -0.189  & 0.013 &  \\
$\psi_V$  & -0.059 & -0.012 & 0.045 & 0.028\\
$m_s/\bar{m}$ & -0.002 & 0.003 & 0.001 & 0.949 & 0.000 \\
\hline
    & $Z^2_G$ & $\psi_P$ & $Z_{q}$  & $Z_s$ & $\psi_V$ &  \\
 
\end{tabular}
\caption{Correlation matrix of the fit shown in tab. 6.} \label{tab_corr_result_final}
\end{table}

The $\eta$-$\eta'$ mixing angle and the $\eta'$ gluonium content are not substantially modified, but  the $\chi^2$ probability is improved with respect to the previous fits. The $\phi - \omega$ mixing angle, $\psi_V$,  is also slightly changed from $(3.17 \pm 0.09)^\circ$ to $(3.32 \pm 0.09)^\circ$ by our new  measurement of BR$(\omega \to \pi^0 \gamma)$, being $\psi_V$ directly related to the ratio $\Gamma(\phi \to \pi^0 \gamma)/\Gamma(\omega \to \pi^0 \gamma)$ (see eq. \ref{eq_phiv}). 

In fig. \ref{fig_contour} we show the 68\% CL contour of the $\eta'$ related  measurements in the $Z^2_G,\psi_P$ plane where the contribution of each measurement to the fit results can be appreciated.  
\FIGURE{
\centering
\includegraphics[width=0.6\textwidth]{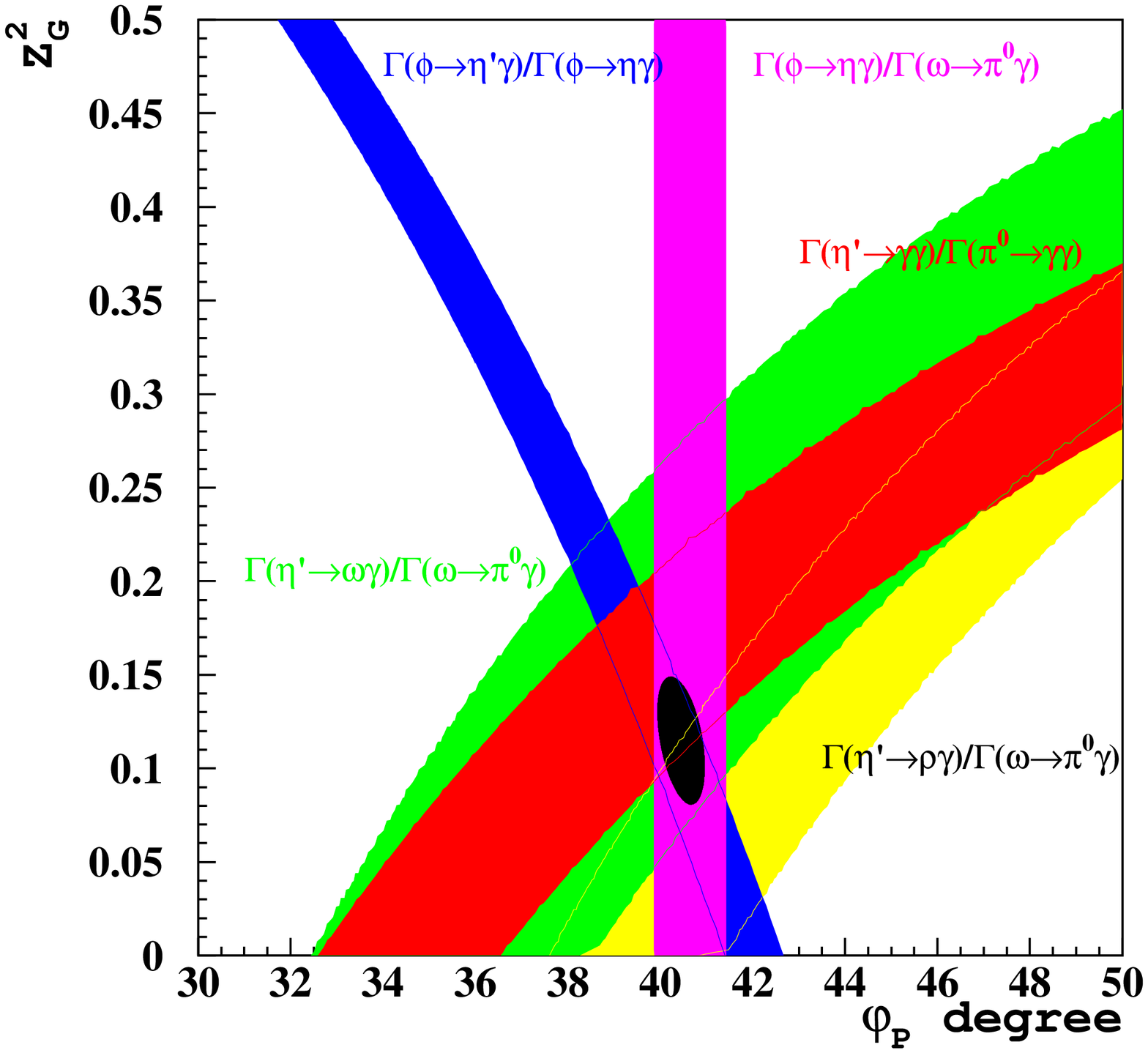}
\caption{68\%  confidence level regions of the shown measurements in the $Z^2_G,\psi_P$ plane.} \label{fig_contour}

}
The best fit values of the width ratios are shown in tab. \ref{tab:ratiofit} together with their correlation coefficients in tab. \ref{tab:widthratcorr}.
\begin{table}
\centering
\begin{tabular}{c|c|c}
width ratio & fitted value  & error \\
\hline
$\frac{\Gamma(\eta' \to \gamma \gamma)}{\Gamma(\pi^0 \to \gamma \gamma)}$ & 570 & 35 \\
$\frac{\Gamma(\eta' \to \rho \gamma)}{\Gamma(\omega \to \pi^0 \gamma)}$ & 0.0735 & 0.007 \\
$\frac{{\Gamma}(\phi \to \eta' \gamma)}{{\Gamma}(\phi \to \eta \gamma)}$ & 0.0047 & 0.0002 \\ 
$\frac{\Gamma(\eta' \to \omega \gamma)}{\Gamma(\omega \to \pi^0 \gamma)}$ & 0.0087 & 0.0010 \\
$\frac{\Gamma(\omega \to \eta \gamma)}{\Gamma(\omega \to \pi^0 \gamma)}$ & 0.0064 & 0.0015 \\
\hline
\end{tabular}
\begin{tabular}{c|c|c}
width ratio & fitted value  & error \\
\hline
 $\frac{\Gamma(\rho \to \eta \gamma)}{\Gamma(\omega \to \pi^0 \gamma)}$ & 0.061 & 0.004 \\
$\frac{\Gamma(\phi \to \eta \gamma)}{\Gamma(\omega \to \pi^0 \gamma)}$ & 0.072 & 0.003 \\
$\frac{\Gamma(\phi \to \pi^0 \gamma)}{\Gamma(\omega \to \pi^0 \gamma)}$ & 0.0079 & 0.0004 \\
$\frac{\Gamma(K^{*+} \to K^+ \gamma)}{\Gamma(K^{*0} \to K^0 \gamma)}$ & 0.43 & 0.06 \\
& & \\
\hline
\end{tabular}
\caption{Fitted values of the $\Gamma$ ratios. } \label{tab:ratiofit}
\end{table}

\begin{table}
\centering
\begin{tabular}{c|cccccccc}
$\eta'\rho\gamma$ & 0.28 &        &   &      &      &   &      & \\ 
$\phi\eta'\gamma$ & 0    &  0     &   &      &      &   &      & \\
$\eta'\omega\gamma$ & 0.24 & 0.48 & 0 &      &      &   &      & \\
$\omega\eta\gamma$ & 0 & 0.04     & 0 & 0.28 &      &   &      & \\
$\rho\eta\gamma$ & 0 & 0.13       & 0 & 0.08 & 0.05 &   &      & \\
$\phi\eta\gamma$ & 0 & 0.28       & 0 & 0.18 & 0.10 & 0 &      & \\
$\phi\pi^0\gamma$ & 0 & 0.17      & 0 & 0.11 & 0.06 & 0 & 0.36 & \\
$K^{*+}K^+\gamma$ & 0 & 0         & 0 & 0    & 0    & 0 & 0    & 0\\
\hline
                  & $\eta'\gamma\gamma$ & $\eta'\rho\gamma$ & $\phi\eta'\gamma$ &$\eta'\omega\gamma$& $\omega\eta\gamma$ &$\rho\eta\gamma$&$\phi\eta\gamma$&$\phi\pi^0\gamma$ \\
\end{tabular}
\caption{Correlation coefficients of the best-fitted values.} \label{tab:widthratcorr}
\end{table}

\section{Prospects with improved measurements}
We have shown that within the precision of the available data the crucial measurement sensitive to the $\eta'$ gluonium content is $\Gamma(\eta' \to \gamma \gamma)/
\Gamma(\pi^0 \to \gamma \gamma)$. 
The theoretical framework used to describe $\eta' \to \gamma \gamma$ and $V \to P \gamma$ transition is  different. In fact while in the first case we have a quark-antiquark annihilation into two photons, in the second case we have a transition among two meson with a photon emission via the spin flip of one of the two quarks.  Therefore it is important both to reach a sensitivity to the gluonium independently from the $\eta' \to \gamma \gamma$ decay and to measure again the $\eta' \to \gamma \gamma$ branching ratio. 

The $\eta' \to \gamma \gamma$ branching ratio affects directly the $\eta'$ decay width while the $\eta' \to \pi^+ \pi^- \eta$ and the $\eta' \to \pi^0 \pi^0 \eta$ branching ratios affect the systematic errors on the $R_{\phi}$ measurement \cite{rphipaper}. The increase of the precision on all main $\eta'$ branching fractions at 1\% level could lead the gluonium contribution to  $\sim$ 4$\sigma$ (see tab. \ref{tab:KLOE2}).  
\begin{table}
 \begin{center}
\begin{tabular}{c|c|c}
& with $\eta' \to \gamma \gamma/\pi^0 \to \gamma \gamma$ & without $\eta' \to \gamma \gamma/\pi^0 \to \gamma \gamma$ \\
\hline
$Z^2_{G}$ & 0.12 $\pm$ 0.03 & 0.11 $\pm$ 0.04 \\
$\psi_P$ & (40.5 $\pm$ 0.6)$^\circ$ & $(40.5 \pm 0.6)^\circ$ \\
$Z_{NS}$ & 0.93 $\pm$ 0.02 & 0.93 $\pm$ 0.03\\
$Z_S$ & 0.83 $\pm$ 0.05 &   0.83 $\pm$ 0.05 \\
$\psi_V$ & (3.32 $\pm$ 0.08)$^\circ$ & $(3.32 \pm 0.09)^\circ$ \\
$m_s/\bar{m}$ & 1.24 $\pm$ 0.07 & 1.24 $\pm$ 0.07 \\
\hline
\end{tabular}
\caption{Fit values assuming 1\% error on the $\eta'$ branching fractions: \textbf{left:} using $\eta' \to \gamma \gamma/\pi^0 \to \gamma \gamma$, \textbf{right:} without using $\eta' \to \gamma \gamma/ \pi^0 \to \gamma \gamma$.} \label{tab:KLOE2}
\end{center}
\end{table}
The results shown in the table are obtained assuming as central value of all measurements  the width ratios shown in tab. \ref{tab:ratiofit}  and assigning to all $\eta'$ branching ratios a  1\% error (the error on $\Gamma_{\eta'}$ and $R_{\phi}$ has been recomputed accordingly); the correlation matrix is taken equal to the present measurements. The fit gives a large $\eta'$ gluonium component even without using the $\eta' \to \gamma \gamma/ \pi^0 \to \gamma \gamma$ measurement (see tab. \ref{tab:KLOE2}, right column).  

\begin{figure}
\includegraphics[width=0.48\textwidth]{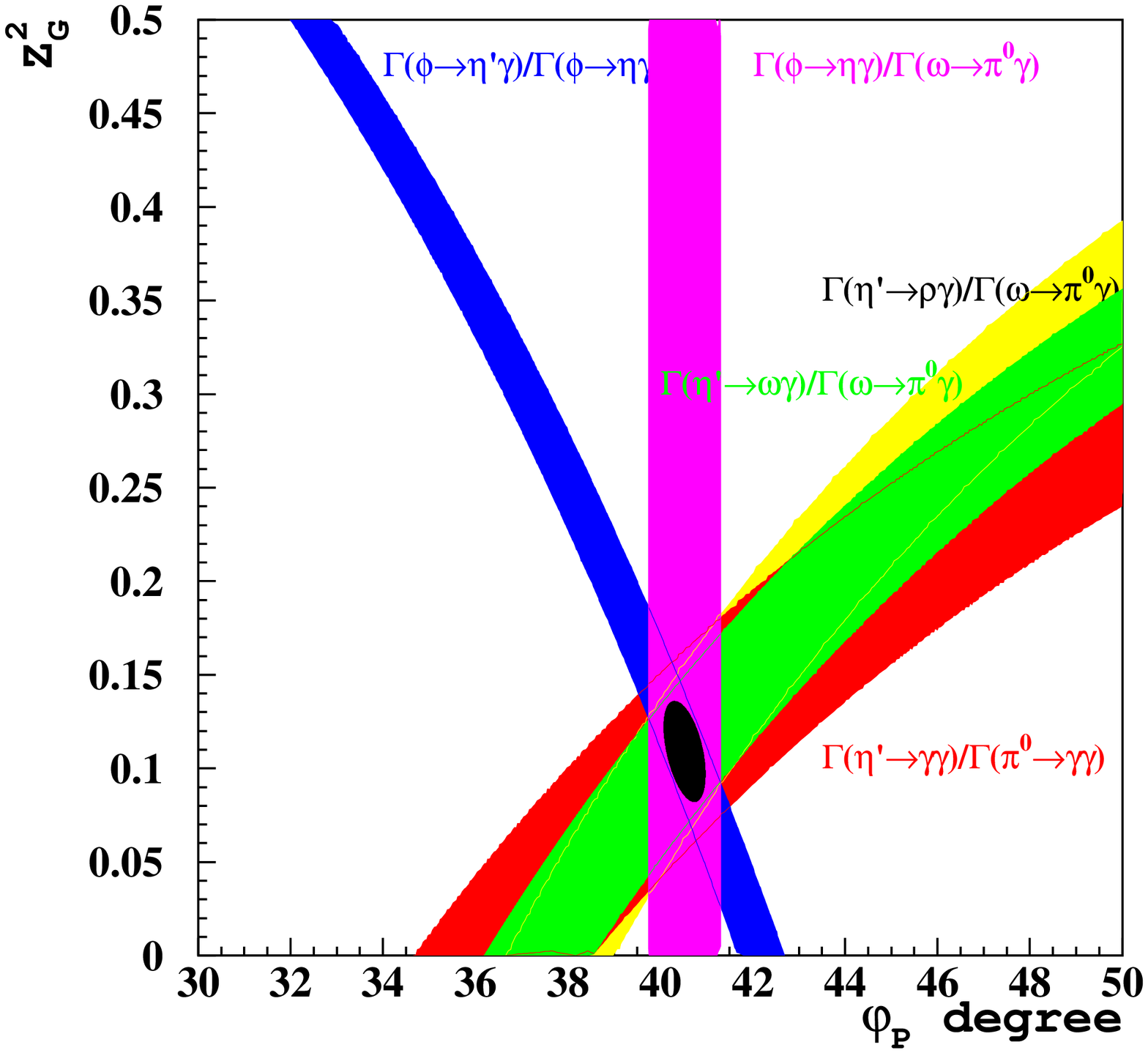}
\includegraphics[width=0.48\textwidth]{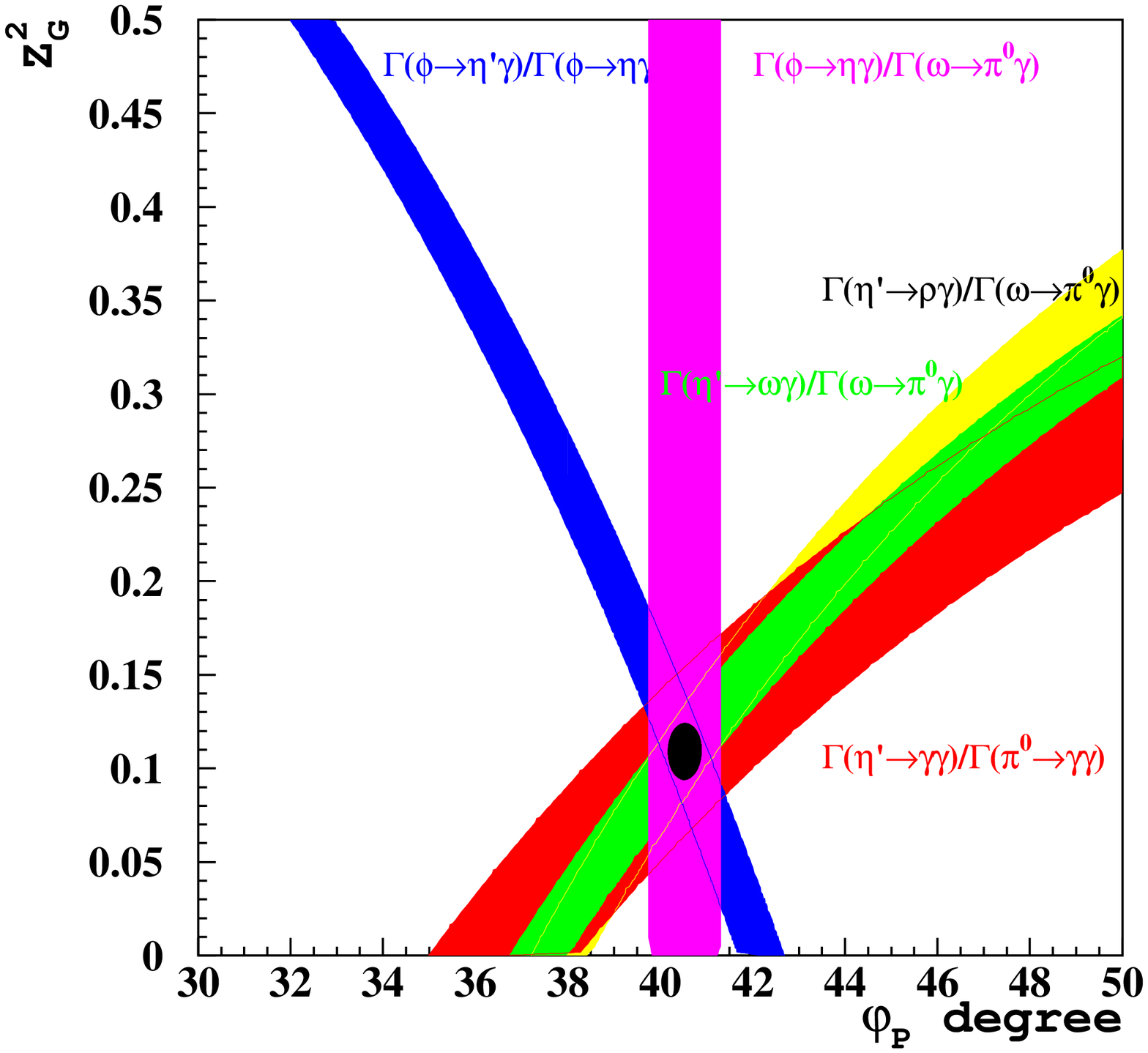}
\begin{center}
\caption{Future prospects: one sigma confidence regions in the $(\psi_P,Z^2_G)$ plane assuming present fitted measurements but with reduced errors: \textbf{left:} with 1\% error on $\eta'$ branching ratios, \textbf{right:} as before plus 1\% precision on $\Gamma(\eta'\to \gamma \gamma)$. } \label{fig:contKLOE2}
\end{center}
\end{figure}

In fig. \ref{fig:contKLOE2} the expected 68\% CL contour in the plane $\psi_P,Z^2_G$ is shown. Notice that the $\eta' \to \gamma \gamma$ measurement would be not  
needed anymore to have a significative gluonium component.

A big improvement in the gluonium content determination can be obtained by measuring  the $\eta'$ decay width directly  through the measurement of $\sigma(e^+ e^- \to e^+ e^- \gamma^* \gamma^* \to e^+ e^- \eta')$. The measurement of this cross section at 1\% level and the $\eta' \to \gamma \gamma$ branching ratio at the same level of accuracy would allow to determine the $\eta'$ width:
\[
\Gamma_{\eta'} = \frac{\Gamma(\eta' \to \gamma \gamma)}{{\rm BR}(\eta' \to \gamma \gamma)}
\]
 with a fractional error of $ \sim 1.4$\%.
In fig. \ref{fig:contKLOE2}, right the 68\% CL contours in the $(\psi_P,Z^2_G)$ parameter plane are shown for all $\eta'$ related measurements. The improvement in the fit accuracy is evident. The gluonium contribution would be measured at $\sim$ 5$\sigma$'s.

\section{Conclusions}
The origin of the difference between the KLOE result \cite{rphipaper} on the $\eta'$ gluonium content  and ref. \cite{Escribano_new}  is the use of the $\Gamma(\eta' \to \gamma \gamma) / \Gamma(\pi^0 \to \gamma \gamma)$ measurement. A global fit to all measured  $V \to P \gamma$ and $P \to V \gamma$ transitions of light mesons has been performed extracting all the relevant parameters. The new results are slightly different from our previous results but confirm the presence of a significative gluonium contribution in the $\eta'$ meson. The origin of this contribution has been investigated and found to stem mainly from the $\Gamma(\eta' \to \gamma \gamma) / \Gamma(\pi^0 \gamma \gamma)$ measurement.
The fit has been updated with all recent measurements from PDG \cite{PDG08} and some disagreement with the PDG fitted value of the BR($\omega \to \eta \gamma$) has been found. 
The  average of the $\omega \to \eta \gamma$ branching ratio is therefore preferred to the PDG fit and a better fit is obtained while
the gluonium content and the mixing angle are unaffected. The result has been updated using the recent KLOE measurement of the $\omega \to \pi^0 \gamma$ branching ratio and and the lattice results for $f_q/f_{\pi}$ and $f_s/f_{\pi}$ assuming exact isospin symmetry. The mixing angle and the gluonium content are again  marginally affected by this measurement while a  different $\phi - \omega$ mixing angle is obtained.

Finally, we have estimated the impact of future improvements in the measurements of the $\eta'$ branching ratios and $\eta'$ width in the gluonium content determination. A 5$\sigma$ evidence could be obtained with a 1\%  precision on each measurement.

\section*{Acknowledgements}
We would like to thank R. Escribano for many useful discussions.
We want to thank our technical staff: G. F. Fortugno and F. Sborzacchi for their dedicated work to ensure an efficient operation of the KLOE computing facilities.
This work was supported in part
by EURODAPHNE, contract FMRX-CT98-0169; 
by the German Federal Ministry of Education and Research (BMBF) contract 06-KA-957; 
by the German Research Foundation (DFG),'Emmy Noether Programme',
contracts DE839/1-4;
and by the EU Integrated
Infrastructure
Initiative HadronPhysics Project under contract number
RII3-CT-2004-506078.

\small{

}

\end{document}